\documentclass[10pt,conference]{IEEEtran}
\usepackage{cite}
\usepackage{amsmath,amssymb,amsfonts}
\usepackage{algorithmic}
\usepackage{graphicx}
\usepackage{textcomp}
\usepackage{xcolor}
\usepackage[pass]{geometry}

\usepackage[hyphens]{url}
\usepackage{fancyhdr}
\usepackage{hyperref}
\usepackage{booktabs}
\usepackage{tabularx}
\usepackage{multirow}
\usepackage{booktabs}   %

\hypersetup{
  colorlinks=false,
  pdfborder={0 0 0}
}

\title{Accelerating Distributed Digital Ising Machines}

\author{
  \IEEEauthorblockN{Tong Wu\IEEEauthorrefmark{1},
    Ziyad Alswaidan\IEEEauthorrefmark{1},
    Shreya Mukherjee\IEEEauthorrefmark{1},
    Ting-Yu Lee\IEEEauthorrefmark{1},
    Daniel Xu\IEEEauthorrefmark{1},\\
    Atharva Raut\IEEEauthorrefmark{1},
    Harry Tye\IEEEauthorrefmark{2},
    Abhinav K. Puthran\IEEEauthorrefmark{1},
    Kemal Selcuk\IEEEauthorrefmark{3},
    Navid A. Aadit\IEEEauthorrefmark{3},\\
    Ken Mai\IEEEauthorrefmark{1},
    Prashanth Mohan\IEEEauthorrefmark{2},
    Kerem Y. Camsari\IEEEauthorrefmark{3},
    Tathagata Srimani\IEEEauthorrefmark{1}}
  \IEEEauthorblockA{\IEEEauthorrefmark{1}Carnegie Mellon University\quad
    \IEEEauthorrefmark{2}University of Pittsburgh\quad
    \IEEEauthorrefmark{3}University of California, Santa Barbara}
}

\begin{document}
\maketitle
\thispagestyle{plain}
\pagestyle{plain}

\newcommand{\placeholder}[1]{\par\medskip\noindent\fbox{\parbox{0.95\linewidth}%
{\footnotesize\textbf{[PLACEHOLDER]}~#1}}\par\medskip}

\newcommand{\TTS}{\ensuremath{\mathrm{TTS}}}

\renewcommand{\topfraction}{0.85}
\renewcommand{\bottomfraction}{0.5}
\renewcommand{\textfraction}{0.15}
\renewcommand{\floatpagefraction}{0.7}
\renewcommand{\dbltopfraction}{0.85}
\renewcommand{\dblfloatpagefraction}{0.7}
\setcounter{topnumber}{2}
\setcounter{bottomnumber}{1}
\setcounter{totalnumber}{3}
\setcounter{dbltopnumber}{2}

\begin{abstract}
Ising machines, dedicated accelerators built around the dynamics of
Ising-like spin models, have emerged as a leading platform for
combinatorial optimization and probabilistic sampling. The problems that
motivate them require millions of spins, far beyond the on-chip capacity
of any single device, so a practical Ising machine must distribute its
computation across many devices. Prior work reports throughput at a
single operating point, leaving the binding resource unidentified and
platform crossovers measurable only by experiment; no calibrated
cross-platform cost model of distributed Ising machines has been reported,
nor have systematic techniques to accelerate their Markov chain Monte Carlo
(MCMC) workloads. This work presents a
calibrated cost model that decomposes digital Ising sweep time into
compute, memory, synchronization, and inter-device communication terms,
validated with 4.8\% mean error across three GPU generations, FPGA and ASIC up to 1 million spins. Guided by the model, we address each term in turn and propose: (i)~a data placement strategy, worth up to 4.2$\times$ speedup; (ii)~multi-rate and
bounded-staleness schedules that relax the barrier that Gibbs sampling
imposes, cutting time-to-solution (TTS) by up to 3.16$\times$ on
three benchmarks; and (iii)~a phase-offset
streaming dataflow that hides boundary exchange behind interior
computation whenever a link satisfies a latency-transparency criterion, yielding up to
1.25$\times$/1.74$\times$/2.01$\times$ on an H200 pair, a ZCU106 pair,
and a four-board XEM8320 $2{\times}2$ mesh. The streaming dataflow is also implemented on a fabricated four-chip 28\,nm ASIC
sustaining 28.1\,Gflips/s at 4.81\,pJ/flip.
\end{abstract}

\section{Introduction}
\label{sec1}

\begin{figure*}[ht]
    \centering
    \includegraphics[width=1\linewidth]{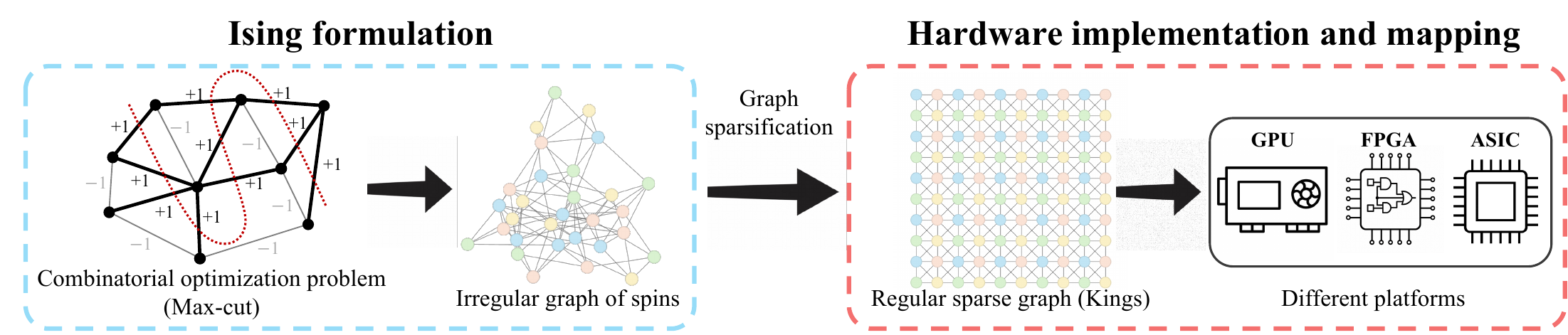}

    \caption{Workflow for mapping a combinatorial optimization problem onto a digital Ising machine. This work focuses on hardware implementation and mapping (red box) across GPU, FPGA, and ASIC platforms.}
    \label{fig1}
\end{figure*}

Combinatorial optimization problems such as logistics routing, resource scheduling, and probabilistic machine-learning inference are ubiquitous, and their underlying NP-hard structure makes them costly to solve at scale~\cite{lucas2014ising}. Probabilistic hardware already reaches several of these classes, training deep Boltzmann networks~\cite{niazi2024training}, running diffusion-based generative models~\cite{singh2026diffusion}, and competing with leading quantum annealers on hard optimization instances~\cite{chowdhury2025pushing}. The Ising model has become a widely used abstraction for such problems: binary variables are represented as \emph{spins}, and pairwise couplings are chosen so that low-energy spin configurations correspond to high-quality solutions~\cite{ising1925beitrag,lucas2014ising,barahona1982computational,mohseni2022ising}. Hardware architectures that exploit this abstraction are broadly called Ising machines.

Ising machines can be realized in many ways. Some implementations embed the dynamics directly in physical substrates, including superconducting quantum annealers with thousands of qubits~\cite{johnson2011dwave,mcgeoch2020dwave,dwave2025advantage2} and optical coherent Ising machines with up to 100,000 spins~\cite{honjo2021cim100k}. Others implement the spin dynamics digitally, including probabilistic-bit machines in which each spin is realized as a stochastic binary unit whose probability is set by its inputs~\cite{chowdhury2023fullstack}. This paper focuses on digital Ising machines~\cite{takemoto2019,yamaoka2017} implemented on GPUs, FPGAs, and ASICs, where the computational kernel is explicit, measurable, and comparable across platforms.

Mapping a problem onto such hardware follows a common workflow: the problem is expressed as an Ising model, its irregular interaction graph is transformed into a sparse, regular, hardware-friendly graph, and the resulting spin-update workload is executed on a target platform (Fig.~\ref{fig1}). Realizing the resulting acceleration at practical problem scales requires millions of spins, which exceed the on-chip capacity of any single digital accelerator~\cite{aadit2026million}. Thus, irrespective of the hardware implementation, scaling ultimately forces the Ising machine to become distributed.
The central question is therefore not simply which device (GPU, FPGA, or ASIC) is faster, but which resource limits scaling on each implementation. One might expect arithmetic to dominate, since each sweep performs local-field accumulation and stochastic spin updates. In practice, it does not. Performance is governed by where couplings reside in the memory hierarchy, whether the working set fits in fast on-chip storage, how often spin updates must synchronize, and how much boundary state must cross device boundaries. These costs depend strongly on graph topology, so the same Ising sweep can be memory-bound on a GPU, synchronization-sensitive under global update barriers, and communication-bound once partitioned across devices.

Our key contributions are as follows:
\begin{itemize}
  \item \textbf{Modeling the distributed Ising sweep} with a calibrated
  four-term decomposition into compute, memory, synchronization, and
  inter-device communication, which reproduces measured runtime to
  2.6--5.2\% mean error across Ada, Blackwell, and Hopper GPUs up to
  million-spin scale, and predicts statically scheduled FPGA and ASIC
  throughput from synthesis-time constants.
 
  \item \textbf{Data-placement techniques that improve the memory term},
  characterizing storage format, residency tier, and datapath precision. Selecting the optimal precision is worth 1.2--4.2$\times$ speedup, yielding a geometric-mean 3.4$\times$ TTS reduction at an identical energy target.
 
  \item \textbf{Multi-rate and bounded-staleness schedules that remove
  the barrier term}, trading per-run success probability for sweep rate, giving 1.75--3.16$\times$ TTS reduction evaluated over MaxCut, frustrated-loop, and
  tile-planted families.
 
  \item \textbf{A phase-offset streaming dataflow that removes the
  communication term}, designing the wave-slot order to
  maximize inter-device slack, so every transfer completes inside a
  latency-hiding window whenever a link satisfies the latency-transparency criterion.
  
  \item \textbf{Evaluation across three substrates}: three classes of GPUs, 1-D and
  2-D Xilinx FPGA meshes, and a fabricated four-chip 28\,nm ASIC~\cite{wu2026pegasus}, where
  the streaming schedule holds per-sweep time flat in chip count at
  0.985\,$\mu$s for 27,648 spins and 4.81\,pJ per update, 
22.1$\times$ faster than a topology-specialized H200 baseline
  and three orders of magnitude lower in energy per
  spin update on the same Pegasus topology.
\end{itemize}

\section{Background}
    \label{sec2}
    \subsection{Motivation}
    \label{subsec:ising_background}
    Ising machines cast computation as the evolution of a dynamical system. Instead of executing a general sequence of instructions, the hardware maintains a network of interacting binary variables and lets their collective dynamics search for low-energy states. This formulation makes the Ising model especially attractive for acceleration: once an optimization problem is expressed as a spin system, the same local interactions that define the energy function also define the structure of the computation. The Ising model, originally formulated to describe ferromagnetism in
    statistical mechanics~\cite{ising1925beitrag}, provides a
    computational framework for a broad class of combinatorial optimization
    problems. The system energy is:
    \begin{equation}
    \label{eq:ising_energy}
        E(m) = -\sum_{i < j}^N J_{ij} m_i m_j - \sum_i^N h_i m_i
    \end{equation}
    where $N$ is the number of spins, $m_i \in \{-1, +1\}$ are binary spin
    variables, $J_{ij}$ are pairwise interaction couplings, $h_i$ are
    external fields, and $m = [m_1, \ldots, m_N]^\top$ denotes the spin
    configuration.  The ground state of $E$ encodes the solution to the
    optimization problem. The couplings define a graph of interconnections.
    
    Lucas~\cite{lucas2014ising} has shown that all 21 of Karp’s original
NP-complete problems~\cite{karp1972reducibility} can be mapped to Ising
formulations. However, most of these formulations lead to dense connectivity,
whereas practical Ising hardware typically exposes fixed sparse connectivity.
Sparsification techniques, either algorithmic~\cite{callahan2026restoring}
or based on directly reducing edge densities~\cite{spielman2008graph,sajeeb2025scalable},
are therefore essential for bringing these mappings into the high-parallelism,
on-chip regime needed for efficient hardware implementation.

\subsection{Gibbs Sampling}

\label{sec:gibbs_sampling}
To sample low-energy configurations, we target the Boltzmann distribution, which concentrates on ground states at thermal equilibrium. We sample it with Markov Chain Monte Carlo (MCMC) methods such as Gibbs sampling~\cite{geman1984stochastic}, which iteratively updates individual spins conditioned on their neighbors.

For spin $i$, the update rule for a bipolar Ising problem is:
\begin{equation}
    I_i = \sum_{j \in \mathcal{N}(i)} J_{ij} m_j + h_i ,
\label{eq:local_field}
\end{equation}
\begin{equation}
    m_i \leftarrow \operatorname{sgn}\left[\tanh(\beta I_i) - u\right],
\label{eq:gibbs_update}
\end{equation}

\noindent where $\mathcal{N}(i)$ is the set of neighbors of spin $i$; $I_i$ is
the \emph{local field}, and $u \sim \text{Uniform}(-1, 1)$ is a random
sample.  Eq.~\eqref{eq:local_field} computes the weighted sum
of neighbor states, and Eq.~\eqref{eq:gibbs_update} applies a
stochastic sigmoid activation to it.

Here $\beta = 1/T$ is the inverse temperature, which controls the sharpness of
the stochastic update. At high temperature (small $\beta$), updates are
nearly random and the chain explores the configuration space freely;
at low temperature (large $\beta$), updates are driven toward the lowest-energy
states. Since the Boltzmann distribution concentrates on ground states only as
$T \!\to\! 0$, sampling typically begins at high temperature to escape local minima and gradually cools
($\beta$ is raised) until the system settles into a ground state.
This process is called \emph{simulated annealing}~\cite{kirkpatrick1983sa}.

The computational cost per spin update is $O(d_i)$, where
$d_i = |\mathcal{N}(i)|$ is the degree of spin $i$.  A \emph{sweep}
is defined as $N$ attempted spin updates (one per spin). Formally, the spins form a coupling graph (CG) with $N$ spins and $M$ edges. For the regular graphs benchmarked in this paper
the degree is uniform across spins, giving an average degree
$\bar{d} = d_i = 2M/N$.

\subsection{Time-to-Solution}
\label{subsec:tts}
Because the update is stochastic, a single run may or may not reach a
good state, so quality is reported as time-to-solution (TTS): the expected
wall-clock time to reach a target-quality solution with probability at
least $p_{\text{target}}$ (99\% in this
work)~\cite{ronnow2014defining,boixo2014evidence,albash2018demonstration,kowalsky2022xorsat}. A run of duration
$t_f$ succeeds with probability $p_{\mathrm{succ}}$; repeating
independently for $R$ iterations succeeds with probability
$1-(1-p_{\mathrm{succ}})^{R}$, so requiring at least $p_{\text{target}}$
gives $R=\ln(1-p_{\text{target}})/\ln(1-p_{\mathrm{succ}})$ iterations of
cost $t_f$ each:
\begin{equation}
\text{TTS}(t_f) = t_f \cdot \frac{\ln(1 - p_{\text{target}})}{\ln(1 - p_{\mathrm{succ}})}.
\label{eq:tts}
\end{equation}
$t_f$ and $p_{\mathrm{succ}}$ trade off against each other,
discussed further in Section~\ref{sec5}.

\subsection{Update Schedules, Sparsification, and Graph Topology}
\label{sec:sync_async}

Gibbs sampling is sequential by construction: each spin update must observe the most recent values of its neighbors to preserve detailed balance, and naively updating all spins in parallel violates this condition~\cite{aarts1989simulated}. Fully asynchronous updates, in which each spin fires at a random instant, relax this requirement and are natural for stochastic devices~\cite{11353558,aadit2022sparse,cook2019335}. The induced statistical error is theoretically bounded, and for sparse problems the iteration count grows by only a factor of $1 + O(1/N)$~\cite{pmlrv48sa16}, making asynchrony attractive when correctness can be traded for throughput.

The principled middle ground is colored Gibbs sampling, which separates spins into $\chi$ independent sets such that no two spins of the same color share an edge~\cite{gonzalez2011parallel,aadit2022sparse}. All spins of one color update simultaneously, and the effective parallelism is $N/\chi$ per color step. The chromatic number $\chi(G)$ thus becomes the central architectural parameter: minimizing $\chi$ maximizes the parallel speedup that any platform can extract from a colored update schedule.

Because $\chi$ is set by graph topology, the choice of graph dictates both parallelism and memory behavior. Dense all-to-all graphs are the worst case on both axes: their $O(N^2)$ coupling matrix exhausts on-chip memory at moderate $N$, and because $\chi = N$ the parallelism collapses to $O(1)$, leaving barrier overhead as the dominant cost. Sparse regular graphs invert both problems. The 8-connected kings graph (Fig.~\ref{fig1} right panel) and 15-connected Pegasus graph serve as running examples throughout this paper; both have $O(N)$ memory footprint, $\chi = 4$, and predictable high-locality access patterns, which suits them to both GPU coalescing and FPGA on-chip routing~\cite{matsubara2020cmos}.

Although the experiments use the regular graphs as controlled benchmarks, the execution model is not limited to regular graphs. Irregular sparse graphs can also be colored and updated in parallel, and many require only a small number of colors \cite{aadit2022sparse}. 
The same model uses an irregular sparse graph's $\bar d$, $M$, $\chi$, and storage format.

Dense mapped instances are a separate regime known to scale poorly in digital Ising hardware~\cite{sajeeb2025scalable}. This paper therefore focuses on sparse scalable hardware graphs.

Real-world optimization problems frequently map to dense graphs, so sparsification~\cite{aadit2022sparse} is what makes hardware-efficient execution possible. Replacing an all-to-all instance with a structured sparse embedding shrinks the working set and reduces the number of floating point operations per spin. 
\leavevmode TTS measures solver execution on the mapped hardware graph and excludes the one-time mapping step. For reusable problem families, a sparse master graph can be embedded once and reused across instances~\cite{nikhar2024xorsat}, while dense-to-sparse mapping cost and quality are characterized separately~\cite{sajeeb2025scalable}.

\leavevmode
Gibbs sampling targets a prescribed Boltzmann distribution, imposing a fundamental constraint on parallelism: directly interacting spins cannot be updated in parallel using independent single-site rules without altering the stationary distribution. As connectivity increases, the available parallelism decreases, collapsing to sequential updates on fully connected graphs. This is not merely an implementation limitation, but an unavoidable consequence of the sampling algorithm itself.

Alternative approaches, such as stochastic cellular automata and momentum annealing, relax these constraints to achieve apparent parallelism. The former permits simultaneous updates of interacting spins, while the latter introduces inertia that suppresses rapid spin changes. However, parallel spin updates do not necessarily translate into faster convergence or time to solution, nor do they establish equivalence to Gibbs sampling. Our work specifically addresses the architectural costs imposed by exact Gibbs dynamics, including synchronization and communication between dependent updates. Extending the model to alternative dynamics would require revisiting these dependencies and their associated costs.

\section{Modeling the Bottleneck}
\label{sec3}

\begin{table}[t]
\centering
\caption{Recurring notation in the model and distributed schedules.}
\label{tab:symbols}
\footnotesize
\setlength{\tabcolsep}{3pt}
\renewcommand{\arraystretch}{1.05}
\begin{tabularx}{\columnwidth}{@{}lX@{}}
\toprule
\textbf{Symbol} & \textbf{Meaning and units} \\
\midrule
$N$, $K$, $N_{\mathrm{dev}}$ & Total spins, device count, and spins per device \\
$d$, $\bar d$ & Spin degree and average graph degree \\
$\chi$, $\varphi_{\mathrm{IT}}$ & Color counts of the coupling graph and interconnect \\
$S$, $B$ & Parallel compute units (SMs or PEs) and threads per block \\
$W$, $\sigma$ & Waves per color step and order of (color, wave) slots \\
$F$, $F_{\mathrm b}$ & CG color steps and scheduled IT phases between synchronizations, respectively \\
$\gamma$, $\gamma_{\mathrm b}$ & Number of within and cross device barriers per sweep \\
$b_J$, $Q_{\mathrm{bit}}$ & Coupling width and memory traffic per update (bits) \\
$\eta_{\mathrm{mem}}$ & Fraction of effective memory bandwidth sustained \\
 $\Theta_K$ & Aggregate spin-update throughput (updates/s) \\
\bottomrule
\end{tabularx}%

\end{table}

Table~\ref{tab:symbols} collects the recurring symbols in this paper.

\begin{figure}[t]
    \centering
    \includegraphics[width=\linewidth]{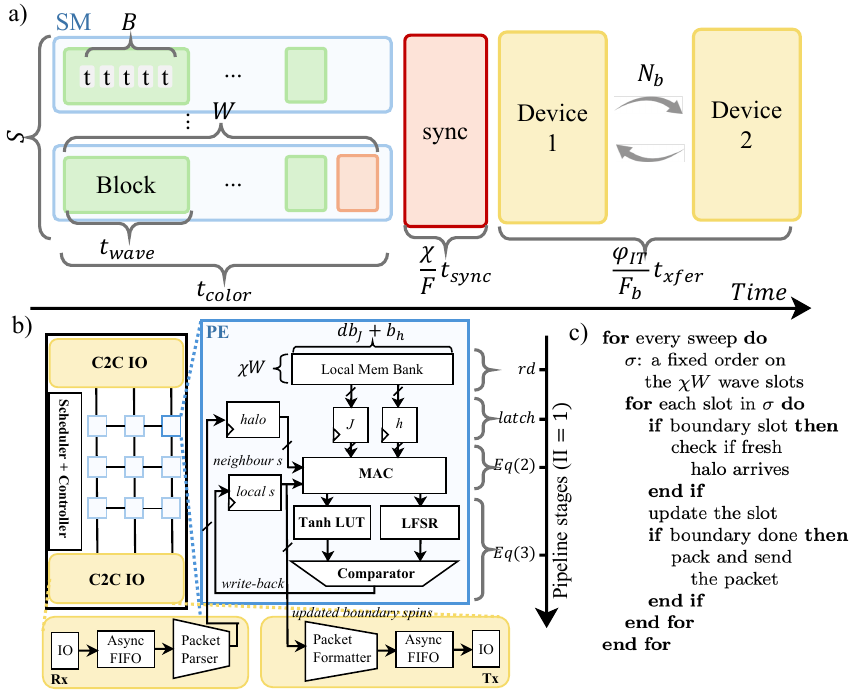}
    \caption[(a) One sweep as Eq.~\eqref{eq:sweep_unified} reads it. A
    straggling block marked in orange. (b) The SP architecture: a processing-element (PE) array under one scheduler, a local memory bank per PE, chip-to-chip (C2C) IO on the edges. The update algorithm is shown in pseudocode.]{(a) One sweep as Eq.~\eqref{eq:sweep_unified} reads it. A
    straggling block marked in orange. (b) The SP architecture: a processing-element (PE) array under one scheduler, a local memory bank per PE, chip-to-chip (C2C) IO on the edges. The update algorithm is shown in pseudocode.}
    \label{fig2}
\end{figure}

Each Ising spin update moves a fixed set of operands through
Eqs.~\eqref{eq:local_field} and~\eqref{eq:gibbs_update}. The machine fetches couplings $J$ and bias $h$, accumulates the local field, draws a random sample, and writes the new spin state. An example datapath is shown in Fig.~\ref{fig2}b.  A sweep applies this update once to every spin, so $N$ updates complete one sweep. We decompose sweep time into compute, memory, synchronization, and inter-device communication costs, then use this decomposition to identify bottlenecks on dynamically scheduled GPUs and statically scheduled streaming platforms (SPs) like FPGA and ASIC. For GPUs, we instantiate the model on a CUDA kernel~\cite{artifact2026} and calibrate and validate each term across the Ada RTX 4070, Blackwell RTX 6000, and Hopper H200. Since statically scheduled platforms are deterministic, they can be modeled from synthesis-time constants alone, with no empirical fit.

\subsection{The Sweep-Time Decomposition}
\label{subsec:unified_model}

We compare platforms using spin-update throughput, $\Theta=N/t_{\mathrm{sweep}}$, measured in attempted updates, or flips per second (Eq.~\eqref{eq:gibbs_update}). This schedule-dependent metric remains comparable across platforms even though the accepted-flip rate falls at low temperature. As Fig.~\ref{fig2}a illustrates, we decompose the time per sweep as:

\begin{equation}
\label{eq:sweep_unified}
t_{\mathrm{sweep}} =
\chi \cdot W \cdot t_{\mathrm{wave}}
  \;+\; \gamma \cdot t_{\mathrm{sync}}
  \;+\; t_{\mathrm{comm}}.
\end{equation} 
Here, $\chi$ is the chromatic number of the CG; $W$ is the number of serialized waves per color step (Section~\ref{subsec:exec_quantization}); $\gamma$ is the number of synchronization points per sweep; $t_{\mathrm{sync}}$ is the cost of each synchronization within a device, and $t_{\mathrm{comm}}$ is the inter-device communication time (Section~\ref{sec5}). This additive form follows the BSP~\cite{valiant1990bsp} and LogP~\cite{culler1993logp}
practice of separating compute, communication, and barrier costs.

This model only predicts $t_{\mathrm{sweep}}$, and for a fixed number of sweeps $n_{\mathrm{sweeps}}$, the run duration in
Eq.~\eqref{eq:tts} is $t_f=n_{\mathrm{sweeps}}\,t_{\mathrm{sweep}}$.

Following the roofline model~\cite{williams2009roofline}, we bound each wave by compute or memory bandwidth: $t_{\mathrm{wave}}=\max(t_{\mathrm{compute}},
t_{\mathrm{mem}})$ (Sections~\ref{subsec:arithmetic_intensity} and~\ref{subsec:memory_term}). On a GPU the coupling fetch stalls the wave (Section~\ref{subsubsec:gpu_regime}). On an SP the fetch occupies its own pipeline stage, so $t_{\mathrm{mem}}$ overlaps $t_{\mathrm{compute}}$ and the initiation interval (II) determines the performance (Section~\ref{subsec:streaming_model}).

Closed forms for each term in Eq.~\eqref{eq:sweep_unified} let us compute platform crossovers from the workload parameters ($N,\bar d,\chi,K$) and a few measured hardware constants, using the released model~\cite{artifact2026}, rather than discovering those crossovers by experiment.

\subsection{Wave Quantization}
\label{subsec:exec_quantization}
Finite execution width quantizes each color step into one or more serialized waves. Each platform updates spins with $S$ compute units in parallel, using SMs on a GPU and PEs on an FPGA or ASIC. GPUs dynamically dispatch blocks of $B$ threads to SMs (Fig.~\ref{fig2}a), whereas an SP statically assigns spins to PEs and has $B=1$. Because exact Gibbs sampling requires a global barrier between color steps, one wave can update at most $S\cdot B$ spins per color step. Thus, $N_{\mathrm{crit}}=\chi B S$ marks the largest problem with no wave serialization. Beyond this threshold, the $N/\chi$ updates in each color step require $W=\lceil
N/N_{\mathrm{crit}}\rceil$ waves, so $t_\mathrm{color} =W\cdot t_{\mathrm{wave}}$. Even one block beyond a multiple of $S$ adds a full $t_{\mathrm{wave}}$ while the remaining units wait (Fig.~\ref{fig2}a). This wave quantization is equivalent to inflating the chromatic number from $\chi$ to $\chi W$. This motivates multi-device scaling. We define a slot as one (color, wave) pair, and all the $\chi W$ slots in a sweep
are scheduled with order $\sigma$ (Fig.~\ref{fig2}b). On a GPU, dynamic block dispatch provides no within-grid ordering guarantee, so global barriers separate colors and enforce a color-major $\sigma$. On an FPGA or ASIC, compile-time dependency resolution permits any legal ordering of the $\chi W$ slots, which helps hide the I/O latency (Section~\ref{sec5}).

\subsection{The Compute Term and Arithmetic Intensity}
\label{subsec:arithmetic_intensity}
Each spin update is a sparse dot product of  $d$  multiply-accumulates, a bias add, and one spin write (Eq.~\eqref{eq:gibbs_update}): $2d+1$ FLOPs plus a degree-independent overhead of one sigmoid and six integer operations since we use xorshift32~\cite{marsaglia2003xorshift}; linear-feedback shift register (LFSR) class pseudorandom number generators (PRNGs) suffice for optimization~\cite{aadit2022sparse,singh2024cmos} while learning may need xoshiro or Philox~\cite{blackman2021scrambled, salmon2011parallel, singh2024cmos}. On SPs the generator is more an architectural than a quality choice. On a 26-spin fully-connected FPGA array, four PRNG configurations and an external true random number generator (TRNG) land within 0.49 points of ground-state success. If the LFSR's repeating period is small, or if one LFSR is shared between PEs, the stream repeats or there could be correlated sequential reuse and the solution quality will degrade. Also notice that a 1-bit spin state ($s_i\in\{0,1\}$) turns each multiply into a conditional add.

Ising sweeps sit well below the roofline ridge and are memory-bandwidth-limited for both sparse and dense graphs. The relevant arithmetic intensity is $\mathrm{AI}=8(2d+1)/[d(b_J+b_s)+b_{\mathrm{fix}}]$, with $b_J$ and $b_s$ the coupling and spin-state widths in bits and $b_{\mathrm{fix}}$ the fixed per-update traffic (the bias read and the
spin write-back, $b_h + b_s$). For moderate $d$, AI approaches $16/(b_J+b_s)$, and even at minimal precision $\mathrm{AI}=1.8$ FLOP/byte. Because both operations and traffic scale with $d$, sparse and dense graphs have similar AI. This metric is less predictive for SP implementations, where distinct pipeline stages execute per-spin operations concurrently and the initiation interval sets the bottleneck (Section~\ref{subsec:streaming_model}) \cite{dasilva2013fpga_roofline}.

The kernel is memory-bound at every measured operating point; on the RTX~4070, ALU utilization reaches only $22\%$ at $d=16$. Its compute work is structurally unbalanced across $O(d)$ floating-point accumulations, one sigmoid on the special-function unit, and a small number of integer operations for random-number generation and addressing. Ada and Hopper execute these three pipelines concurrently, whereas Blackwell time-multiplexes them on a unified datapath.

\subsection{The Barrier Cost}
\label{subsec:coord_cost}
The barrier term $\gamma\,t_{\mathrm{sync}}$ and
the exchange term $t_{\mathrm{comm}}$ in Eq.~\eqref{eq:sweep_unified} can be interpreted as one coordination cost at both within-device and across-device scales. 
Colored Gibbs sampling forces a global barrier between color classes so each spin sees committed neighbor values. On GPUs, $\mu$s-scale synchronization makes $t_{\mathrm{sync}}$ significant at small $B$ (Fig.~\ref{fig3}a). 
We parameterize synchronization
frequency by the \emph{synchronization interval} $F$: the number of
color steps between global barriers, giving $\gamma = \chi/F$
synchronization points per sweep.
$F = 1$ gives exact Gibbs where we synchronize after every color step while $1<F < \infty$ gives bounded staleness discussed in Section~\ref{subsec:exact_sched}.

Raising $F$ shrinks the barrier term $\gamma\, t_{\mathrm{sync}}$ but
degrades success probability $p_{\mathrm{succ}}(\gamma)$, since spins
observe increasingly stale neighbor states; the opposing effects
produce a concave TTS speedup curve shown in
Fig.~\ref{fig7}.

GPU barrier cost was measured by injecting redundant barriers into the cooperative kernel;
the slope of sweep time versus barrier count gives $t_{\mathrm{sync}}$ (Fig.~\ref{fig3}a). The barrier
cost is a function of participating blocks, and
RTX~6000 Blackwell has the largest $t_{\mathrm{sync}}$ because its cooperative grids are largest. Barriers are also not pure overhead: removing them at
$N=2\times10^6$ on the RTX~4070 makes the run $42\%$ \emph{slower}, since threads drift out of lockstep and lose streaming
locality on a bandwidth-bound device, as illustrated by the grid-striding transition in Fig.~\ref{fig3}a. Thus, the barrier acts as a cache-locality synchronizer beneficial to bounded staleness discussed in Section~\ref{sec5}.

On SPs, whose pipeline-control barriers propagate in nanoseconds, $t_{\mathrm{sync}} = \mathcal{O}(\text{ns})$; compile-time work assignment gives deterministic timing. Color-step synchronization comes from pipeline control where the only cost is the \emph{pipeline drain},
bounding in-flight writes of the finishing color before the next color reads them.

\subsection{The Memory Term}
\label{subsec:memory_term}

On a platform that fetches couplings through a memory hierarchy, $t_{\mathrm{compute}}$ does not bind, so
$t_{\mathrm{wave}}$ is the memory service time and $t_{\mathrm{mem}}$ sets single-device throughput. Layout and
residency tier are design choices (Section~\ref{sec4}); here we model the traffic each update
moves and the service time it implies.

\subsubsection{Memory traffic}
\label{subsubsec:mem_traffic}

The memory is modeled with a hierarchy-traversal service time and a residency-weighted, occupancy-dependent service rate rather than the fixed peak bandwidth that roofline bounds by. The working set picks the bandwidth tier; the traffic sets how long that tier is occupied, and traffic always exceeds memory footprint. Per sweep, $J$, indices, and $h$ are \emph{streaming} (one read, no reuse),
while each spin state is read on average $d$ times; with spin cache hit rate $\rho_s\in[0,1]$, the per-spin
traffic is
\begin{equation}
\label{eq:Qbit}
Q_{\mathrm{bit}} =
  \underbrace{d\,(b_{\mathrm{idx}} + b_J)+b_{h}}_{\substack{\text{coupling, self-bias}\\ \text{(streaming, no reuse)}}}
  + \underbrace{(1 - \rho_s)\,d\,b_s}_{\substack{\text{neighbor spins}\\ \text{(used $d\times$)}}} \,
  + \underbrace{b_s}_{\substack{\text{spin}\\ \text{write-back}}},
\end{equation}
where $b_{\mathrm{idx}}$ is the column-index width and $b_h$ the bias width, and the per-sweep total is $N\cdot Q_{\mathrm{bit}}$. On SPs with private per-PE memory, $\rho_s\to1$ by
construction. On GPUs, $\rho_s$ comes from L1/L2 hit
counters. The memory traffic is
streaming-dominant: zero-reuse fetches dwarf the spin terms, favoring memory-throughput-optimized
architectures over cache-based ones, and under GPU LRU eviction the zero-reuse streams can even evict spin data
(Section~\ref{subsec:gpu_memory}).

\subsubsection{GPU memory service time}
\label{subsubsec:gpu_regime}

On GPUs, the low arithmetic intensity reduces $t_{\mathrm{wave}}$ to the memory service time:

\begin{equation}
\label{eq:twave_gpu}
\begin{aligned}
t_{\mathrm{wave}} &= \frac{S \cdot B \cdot Q_{\mathrm{bit}}}
         {\eta_{\mathrm{mem}} \cdot \mathrm{BW}_{\mathrm{max}}}, \\[2pt]
\eta_{\mathrm{mem}}(x) &= \frac{\mathrm{BW}_{\mathrm{achieved}}(x)}
         {\mathrm{BW}_{\mathrm{max}}}
  = \eta_{\max}\cdot f(x),
\end{aligned}
\end{equation}
where $B\cdot Q_{\mathrm{bit}}$ is one block's traffic, $\mathrm{BW}_{\mathrm{max}}$ the effective device
bandwidth (a three-level harmonic blend of measured L1/L2/DRAM bandwidths weighted by hit rate), and $\eta_{\mathrm{mem}}\in(0,1]$ the fraction of that bandwidth sustained,
below 1 when memory-level parallelism is too low to saturate the pipeline. $f(x)$ is a saturating hyperbola as a function of the resident-thread pool $x$: by Little's Law~\cite{little1961queuing}, delivered bandwidth grows linearly with the requests in flight and levels off at the ceiling once enough threads are resident to cover the load latency~\cite{volkov2016latency}.

\subsubsection{Statically scheduled platforms}
\label{subsec:streaming_model}

In SPs, $t_{\mathrm{wave}}$ in Eq.~\eqref{eq:sweep_unified} is set by the pipeline's \emph{initiation interval}
$\mathrm{II}$, giving $\Theta=S f/\mathrm{II}$ and
\begin{equation}
\label{eq:sweep_streaming}
t_{\mathrm{sweep}} = \frac{N \cdot \mathrm{II}}{S \cdot f}
  + \gamma \cdot t_{\mathrm{drain}},
\end{equation}
where $\mathrm{II}$ is fixed at synthesis by the most-contended resource (DSP block, memory port, or write-back
path) and $f$ the clock frequency. Unlike a GPU's fixed-in-silicon $S$, an SP sets $S$ at synthesis, coupling
the free parameters: added PEs draw on a shared budget of memory ports, LUTs, and annealing DSPs, and past a
point exhaust it or force a lower $f$ or higher $\mathrm{II}$ to close timing.

\subsubsection{The throughput ceiling}
\label{subsubsec:modeling_insight}

On SPs the unified model collapses to a closed-form ceiling on $\Theta$. With proper pipelining, $t_{\mathrm{drain}}\to0$
(Section~\ref{subsec:coord_cost}) and $\eta_{\mathrm{ovlp}}\to1$
so $t_{\mathrm{comm}}\to0$ (Section~\ref{subsec:comm_term}), giving $t_{\mathrm{sweep}}=\chi W t_{\mathrm{wave}}$.
Expanding $W$ and $t_{\mathrm{wave}}$ at large $N$ ($\lceil N/N_{\mathrm{crit}}\rceil\cdot
N_{\mathrm{crit}}\approx N$),
\begin{equation}
\label{eq:throughput_ceiling}
\Theta_{\mathrm{SP}}
= \frac{N}{t_{\mathrm{sweep}}}
= \min\,\!\Biggl(
  \underbrace{\frac{S \cdot f}{\mathrm{II}}}_{\substack{\text{compute ceiling}\\ \text{(on-chip)}}},\;
  \underbrace{\frac{\eta_{\mathrm{mem}}\,\mathrm{BW}_{\mathrm{max}}}{Q_{\mathrm{bit}}}}_{\substack{\text{bandwidth ceiling}\\ \text{(off-chip)}}}
  \Biggr).
\end{equation}

The terms are equal when the working set is on-chip, since per-PE memory is provisioned to feed the pipeline;
the residency threshold (Eq.~\eqref{eq:l2_threshold}) marks the cliff between regimes, and both ceilings are
\emph{independent of $N$} -- within a regime the streaming machine delivers constant throughput and per-sweep
time exactly linear in $N$. The same reduction for the GPU (Eq.~\eqref{eq:twave_gpu}) sends
$\Theta_{\mathrm{GPU}}\to\eta_{\mathrm{mem}}\, \mathrm{BW}_{\mathrm{max}}/Q_{\mathrm{bit}}$ as $N\to\infty$, the
\emph{same} bandwidth ceiling as $\Theta_{\mathrm{SP}}$, reached only after amortizing the barrier tax and
climbing the occupancy ramp $\eta_{\mathrm{mem}}<1$. At large $N$ the levers are $\mathrm{BW}_{\mathrm{max}}$,
$\eta_{\mathrm{mem}}$, and $Q_{\mathrm{bit}}$ -- why the degree cancellation of
Section~\ref{subsec:arithmetic_intensity} does not imply degree-independent throughput.

\subsection{Model Calibration}
\label{subsec:benchmarks}

\begin{figure}[t]
    \centering
    \setlength{\abovecaptionskip}{2pt}
    \includegraphics[width=\linewidth]{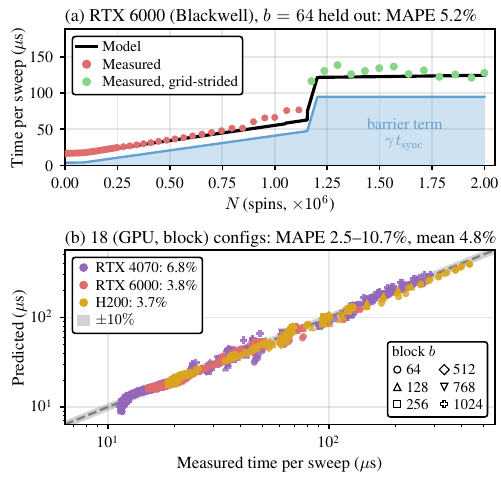}
    \caption[Evaluation provenance]{One calibration per GPU (the $B{=}256$ curve and a measured barrier microbenchmark) predicts
    sweep time at held-out block sizes with no refitting. (a) Exemplar sweep on the RTX~6000 at a
    block size the calibration never sees, with the modeled barrier term shaded and the grid-striding
    sizes marked. (b) Model against measurement for all 18 (GPU, block) configurations, 63 sizes
    each; the band is $\pm10\%$.}
    \label{fig3}
\end{figure}

Block size $B$ packages the $N/\chi$ spins per color step into blocks across the $S$ SMs, trading occupancy
against wave quantization. The grid-wide barrier requires cooperative launch, which caps the grid at the hardware's resident-thread limit; past $N_{\mathrm{crit}}$ each thread loops over multiple spins (a grid-stride loop), and $t_{\mathrm{sweep}}$ jumps 50\% in one step (Figure~\ref{fig3}a).  We fixed $B{=}256$ for
calibration, which predicts held-out sweeps (Fig.~\ref{fig3}b) to $2.5$--$10.7\%$ mean absolute percentage error (MAPE), mean $4.8\%$ over 18 (GPU, block) configurations for $B{=}64$--$1024$.

The model takes architectural constants from the datasheet ($S$, DRAM bandwidth, cache capacity) and a few
empirical parameters measured once per (GPU, kernel) pair~\cite{artifact2026}: cache hit rates from \texttt{ncu}
feed the harmonic blend for $\mathrm{BW}_{\mathrm{max}}$, a $t_{\mathrm{sync}}$ microbenchmark
(Section~\ref{subsec:coord_cost}) supplies the barrier term, and a least-squares fit of the $B=256$ sweep
gives the three constants of $\eta_{\mathrm{mem}}(x)$. Across the full
sweep ($N{=}100$ to $2\times10^6$) on the RTX~4070, RTX~6000, and H200, the calibrated model reaches MAPE
$5.2\%$/$2.6\%$/$3.3\%$ at the calibration block size and $7.2\%$/$4.1\%$/$3.8\%$ averaged over five held-out
block sizes, as summarized in Fig.~\ref{fig3}b. Refitting on half the problem sizes and testing on the disjoint half moves error by at most one point on every GPU, so platform dependence is
confined to a few physically grounded constants, not the decomposition's form of Eq.~\eqref{eq:sweep_unified}; supplying the exact measured
$\eta_{\mathrm{mem}}(x)$ in place of the three-constant fit tightens the match to $\mathrm{MAPE}<2\%$ on all
three GPUs, proving the residual is the price of summarizing $\eta_{\mathrm{mem}}$ in closed form.

With the bottlenecks localized, the rest of the paper removes these terms one at a
time: $t_{\mathrm{mem}}$ in Section~\ref{sec4}, and $t_{\mathrm{sync}}$ and $t_{\mathrm{comm}}$ in Section~\ref{sec5}.

\section{Data Placement: $t_\mathrm{mem}$}
\label{sec4}
\label{sec:memterm}

As $N$ increases, workloads spill off chip. The dominant cost differs across platforms: on a cache-resident GPU at $N{=}10^4$, $t_{\mathrm{mem}}$ and $\gamma t_{\mathrm{sync}}$ account for $79\%$ and $20\%$ of sweep time respectively, while $t_{\mathrm{mem}}$ rises to $96\%$ after spilling at $N{=}10^6$. On the ZCU106, $t_{\mathrm{compute}}$ accounts for $92\%$ with $J$ in BRAM and $77\%$ when streamed from DDR. Therefore, it is crucial to optimize for $t_{\mathrm{mem}}$, which is discussed in this section.

\subsection{Storage Format and Footprint}
\label{subsec:storage}

\begin{figure}[tbp]
    \centering
    \setlength{\abovecaptionskip}{2pt}
    \includegraphics[width=\linewidth]{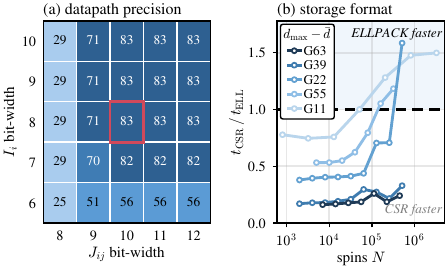}
    \caption[Evaluation provenance]{Coupling precision and storage format trade-offs.
(a) Measured ground-state success rate (\%) on a 26-spin, fully connected array as a function of integer field width and stored coupling width (averaged over 100 Sherrington-Kirkpatrick instances, evaluated at 3 fractional bits). The boxed cell highlights the optimal design point.
(b) Measured execution time ratio ($t_{\mathrm{CSR}}/t_{\mathrm{ELL}}$) on an NVIDIA RTX 4070 against spin count $N$, five G-set instances block-replicated from their published sizes, ordered by irregularity. The three near-regular instances favor ELLPACK as $N$ grows.}
    \label{fig4}
\end{figure}

The working set is the coupling matrix $J$, the bias vector $h$, and the spin
state. Dense storage grows as $N^2$ and loses to sparse storage once
$\bar{d}$ falls. Of the two
well-studied sparse formats~\cite{bell2009spmv,filippone2017spmv},
compressed sparse row (CSR) stores a row pointer per spin and packs each
coupling list tightly, while ELLPACK~\cite{rice1985ellpack} drops the
pointers and pads every row to $d_{\max}$. ELLPACK beats CSR when
$d_{\max} - \bar{d} \;<\; b_{\mathrm{ptr}}/({b_{\mathrm{idx}} + b_J})= 0.67$ at our 32-bit pointers/indices and 16-bit couplings with $b_{\mathrm{ptr}}$ the row-pointer width. Thus, ELLPACK only wins on near-uniform-degree graphs. Kings and Pegasus are both regular, so ELLPACK was used in our study. 

On SPs the static schedule makes pointers and indices implicit: only
couplings and bias are stored, in fixed-point Q$m.n$ format
($b_J = m+n+1$ bits with sign). Topologies having higher degree like Pegasus
($d=15$~\cite{boothby2020pegasus,mcgeoch2020dwave}) or annealing schedules with larger $\beta$ need extra integer bits for the larger local field $I_i$ (Eq~\ref{eq:local_field}). Fig.~\ref{fig4}a sweeps the two widths independently. By choosing the optimal precision, we can achieve a geometric-mean 3.4$\times$ TTS reduction at an identical energy target.

On-chip working-set residency is the single most predictive indicator of
performance for this memory-bound workload. For ELLPACK on a regular graph
it requires
\begin{equation}
\label{eq:l2_threshold}
N \cdot \bar{d} \;<\; \frac{\mathcal{C}_{\mathrm{on-chip}}}{b_{\mathrm{idx}} + b_J},
\end{equation}
where $\mathcal{C}_{\mathrm{on-chip}}$ is the on-chip cache capacity. A
mapping meets the \emph{on-chip residency criterion} when
Eq.~\eqref{eq:l2_threshold} holds.

\subsection{GPU Memory Measurement}
\label{subsec:gpu_memory}

A GPU cannot satisfy the on-chip residency criterion at the sizes that
motivate this work. On the
RTX~4070 (128\,KB L1, 36\,MB L2), effective bandwidth drops $2.9\times$ once
$t_{\mathrm{mem}}$ contributes $96\%$ of the sweep past the crossover of Eq.~\eqref{eq:l2_threshold}. On kings
graphs, couplings, indices, and biases carry ${\sim}75\%$ of traffic with
zero reuse, so marking those loads evict-first avoids cache pollution and
yields $1.44\times$ beyond L2. Scattered neighbor access of every spin keeps L1 hit rates
graph-dependent, ${\sim}40\%$ for kings, $13\%$ for Pegasus, since kings has a better spatial locality. The
near-memory SPs never pay this price of sparse
gather~\cite{sundararaman2024sachi,chiang2024reaim,yang2024sophie}.

CSR walks one spin's coupling list per thread; ELLPACK's column-major
layout lets a warp read 32 consecutive addresses at the cost of an index
multiply-add. CSR's cheaper addressing wins while the kernel is
memory-latency-bound; once memory-bandwidth-bound, ELLPACK's coalesced loads and smaller
footprint win by up to $1.6\times$ (Fig.~\ref{fig4}b). 
The format choice is therefore decided by graph size, degree and irregularity.

\subsection{Streaming Datapath}
\label{subsec:zcu106}

A statically scheduled SP removes
$t_{\mathrm{mem}}$ by holding the working set on chip and feeding every PE from a private local memory. We implement the streaming architecture on a Xilinx ZCU106 and XEM8320 with both kings and Pegasus
graphs in Q6.3 fixed-point, running at 100\,MHz. For kings/Pegasus graphs, every spin's eight/fifteen couplings
and bias form a 90-bit/160-bit BRAM word; updates flow through a three-stage
pipeline at $\mathrm{II}=1$: issue memory read, latch the memory output and
sample the new
state. Here, $t_{\mathrm{drain}}=0$, $t_\mathrm{sync}=0$, and each PE time-multiplexes into
waves once $N/\chi>S$, utilizing $\chi W $ rows of memory. DSP48E2 pre-multipliers were used to implement annealing. 

As $N$ grows, the working set spills through BRAM, LUT-RAM, URAM, and
off-chip DRAM (Fig.~\ref{fig9}, FPGA curves); a prefetch engine on
the DRAM tier cuts the per-spin update time from ${\sim}335$ to
${\sim}3.5$\,ns. Unlike the GPU model, this streaming model needs no
empirical fitting, since its inputs are synthesis-time and datasheet constants and it tracks measured throughput to within a few percent across all
four memory tiers.

\section{Multi-Rate and Bounded Staleness}
\label{sec5}
This section and the next propose methods to remove the synchronization costs $t_{\mathrm{sync}}$ and $t_{\mathrm{comm}}$ of Eq.~\eqref{eq:sweep_unified}, validated on H200 pairs and Xilinx FPGAs.

\subsection{Graph Partitioning}
\label{subsec:comm_term}

A single device has only $t_{\mathrm{sync}}$ and no $t_{\mathrm{comm}}$. As the working set grows, the sweep time on one device becomes marginally more expensive. When the sweep time reduction beats the marginal communication cost, 
partitioning becomes beneficial (Fig.~\ref{fig6}). The coupling graph (CG) is partitioned across $K$ devices physically connected with an
\emph{interconnect topology} (IT). When the IT matches the CG structure, every boundary exchange is single-hop: logical nearest-neighbor couplings become physical links. After partitioning, the per-device spin count becomes $N/K$ and the working set may
fall into the faster memory tier. Also the per-device wave $W(K)$
drops from $\lceil N/N_{\mathrm{crit}}\rceil$ to $\lceil N/(K N_{\mathrm{crit}})\rceil$, but the couplings crossing a partition boundary require periodic state exchange. 

Let $N_{\mathrm{b}}$ be the per-device boundary volume exchanged at each boundary synchronization point, which need not coincide with interior barriers. A balanced 2D partition of a geometrically local, bounded-degree graph (kings shown in Fig.~\ref{fig5}) puts cut
edges on the perimeter, $N_{\mathrm{b}}=O(\sqrt{N/K})$, so the boundary-to-interior ratio vanishes as $N$
grows. The per-sweep
communication time is defined as:

\begin{equation}
\label{eq:comm_unified}
t_{\mathrm{comm}}=\gamma_{\mathrm b}\,t_{\mathrm{xfer}}(1-\eta_{\mathrm{ovlp}}),\;
t_{\mathrm{xfer}}=t_{\mathrm{lat}}+
  \frac{\rho\,N_{\mathrm b}\,b_s}{\mathrm{BW}_{\mathrm{link}}}.
\end{equation}
where $\eta_{\mathrm{ovlp}}$ is the fraction of $t_{\mathrm{xfer}}$ hidden behind
computation; analogously to $\gamma$ in
Section~\ref{subsec:coord_cost} and Eq.~\eqref{eq:sweep_streaming},
$\gamma_{\mathrm b}$ is the number of boundary-exchange rounds per sweep
required by spin dependencies crossing chip boundaries (CBs).
Boundary spins may be exchanged at a lower frequency than the interior
barrier rate $\gamma$. A sweep performs
$\gamma_{\mathrm b}=\varphi_{\mathrm{IT}}/F_{\mathrm b}$ boundary exchanges,
where $\varphi_{\mathrm{IT}}$ is the chromatic number of the IT, the
network-level counterpart of $\chi$ on the CG, and $F_{\mathrm b}$ is the
boundary synchronization interval. $t_{\mathrm{xfer}}$ splits into a latency
term and a bandwidth term~\cite{culler1993logp}. $\rho$ captures the additional
multi-hop link traffic. For matched IT--CG topologies,
$\chi=\varphi_{\mathrm{IT}}$ and $\rho=1$.
For IT--CG mismatches, use the worst-case latency for $t_{\mathrm{lat}}$,
including any additional packet or router overhead. This overhead is
minimized when forwarded data are piggybacked in single-hop packets without
adding a communication round.

We validate this using a King's CG and a
2D-mesh IT on $K=4$ XEM8320s with $N_{\mathrm{dev}}\approx10^6$ and approximately 1k spins per chip edge, giving $N_{\mathrm b}\approx2\,\mathrm{k}$.
Using a SYZYGY link, we have
$t_{\mathrm{lat}} = 341$~ns and
$\mathrm{BW}_{\mathrm{link}} = 0.538$~Gb/s.
Eq.~(11) predicts $t_{\mathrm{comm}}$ within 0.01\% of device measurements.
On $K=2$ H200 GPUs, we have
$t_{\mathrm{lat}} \approx 15.2$~$\mu$s and
$\mathrm{BW}_{\mathrm{link}} \approx 137$~GB/s, with a full helout error of 6.46\%~\cite{artifact2026}. The error is dominated by short boundary-exchange sequences and CUDA event timing settings.

\subsection{Algorithm}
\label{subsec:exact_sched}
\begin{figure}[t]
    \centering
    \includegraphics[width=\linewidth]{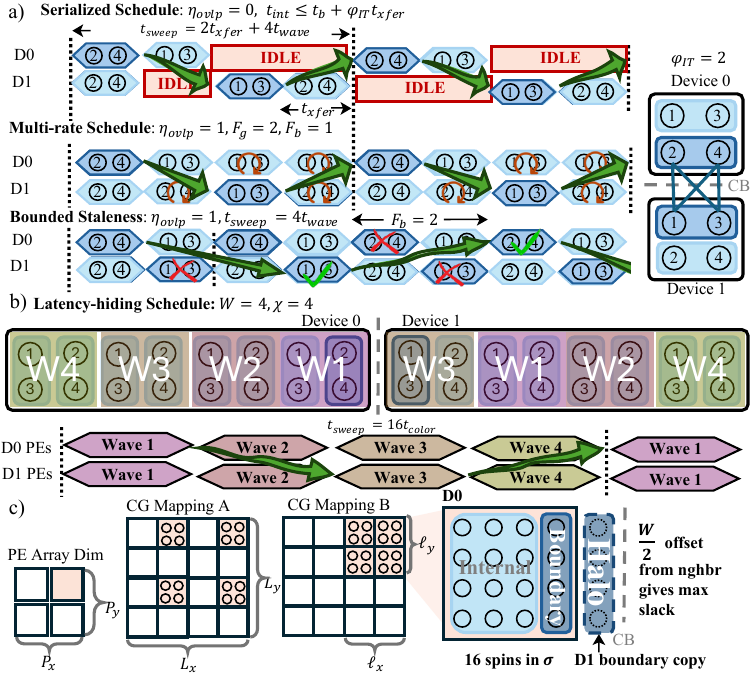}
    \caption{
    Boundary exchange across a two-device cut (boundary spins marked in dark blue). (a) three schedules' timeline: serialized, multi-rate, and
    bounded staleness. (b) Latency hiding
  timeline at $S=1$, $W=4$, $\chi=4$. (c) How the PE array maps to CG with $S=4$, $W=16$, $\chi=4$.}
    \label{fig5}
\end{figure}

Every device runs $\chi W$ wave slots per sweep
(Section~\ref{subsec:exec_quantization}). Fig.~\ref{fig5}a shows an 8-spin kings graph example with each device hosting 4 spins and 1 PE. 
The serialized schedule of Fig.~\ref{fig5}a is the simplest choice, but both
transfers within a sweep ($F_{\mathrm{b}}=1$) land on
the critical path ($\eta_{\mathrm{ovlp}}=0$, $\gamma_{\mathrm{b}}=2$).

 We also observe that a half-duplex link suffices.

The ``IDLE'' periods in the serialized schedule motivate two schedules
that fill them: either update the interior spins (light blue) more
often (multi-rate), or update all spins regardless of whether the
neighbor values have arrived (bounded-staleness) shown in Fig.~\ref{fig5}. 

The multi-rate schedule, which has non-homogeneous spin update rates, is a valid permutation of the colored-Gibbs schedule, and $t_{\mathrm{sweep}}$ is set by
the slowest updating spins. Gating the whole device on the boundary exchange would fold
that wait into every spin's update rate, so we gate per spin rather than per color: a
device without a fresh halo holds only its cut-facing spins and keeps
resampling interior spins. The boundary and interior spins each refresh
with throughputs
$\Theta_{\mathrm{b}}$ and $\Theta_{\mathrm{interior}}$, with the condition of $\Theta_{\mathrm{b}}\, b_s \;\le\; \mathrm{BW}_{\mathrm{link}}$, the sustained link rate. We then define $F_{\mathrm{g}} = \Theta_{\mathrm{interior}}/\Theta_{\mathrm{b}}$, the number of interior updates between two boundary updates (similar to $F_\mathrm{b}$ and $F$, $F_{\mathrm{g}}=1$ is the exact schedule). On a geometrically local graph the interior-to-boundary ratio $N_{\mathrm{int}}/N_{\mathrm{b}}$, with $N_{\mathrm{int}} = N_{\mathrm{dev}} - N_{\mathrm{b}}$ the interior spins of a device holding $N_{\mathrm{dev}}=N/K$ spins, grows as $\sqrt{N_{\mathrm{dev}}}$, so almost all update work runs at full
rate regardless of the link.

For multi-rate scheduling, let $\pi$ be the target
Boltzmann distribution and $P_i$ the Gibbs update kernel for spin $i$.
Since every Gibbs update preserves $\pi$, we have $\pi P_i=\pi$.
If spins are updated at different rates $p_i$, with $p_i>0$ and
$\sum_i p_i=1$, then: $\pi P=\pi\sum_i p_iP_i=\sum_i p_i\pi=\pi$. Thus, provided scheduling is independent of spin values, every spin continues
to update, and current neighbor values are used, multi-rate scheduling changes
mixing times but not the equilibrium distribution.
At $F_{\mathrm{g}}=16$, the boundary spins' integrated autocorrelation time $\tau$ rises about $9\times$ over the exact baseline.

The penalty stays at the boundary. The measured global $\tau$ is unchanged for boundary fractions $N_{\mathrm{b}}/N_{\mathrm{dev}}$ up to $6\%$, rises $3.5\times$ at $25\%$,
and only a $100\%$ boundary fraction slows the machine as a whole (Fig.~\ref{fig7}b).

For bounded staleness, spins use
outdated neighbor information, somewhat reminiscent of a cluster mean-field
approximation, potentially weakening important inter-cluster correlations.
Detailed balance and exact Boltzmann sampling are no longer guaranteed.
The same tradeoff between boundary-exchange rate and accuracy is observed on a
networked million p-bit machine~\cite{aadit2026million}.
Multi-rate
therefore pays its hold in mixing time at zero bias; bounded staleness
pays in solution quality at full $\Theta_\mathrm{b}$, a controlled trade
in the continuum between fully synchronous (correct but slow) and fully
asynchronous (fast but potentially incorrect)
execution~\cite{recht2011hogwild}.

 A boundary spin's staleness, the number of
iterations since its neighbors were last committed, is exactly one under strict Gibbs. Raising $F_{\mathrm{b}}$ cuts bandwidth
demand even when $\eta_{\mathrm{ovlp}}$ reaches 1 (Fig.~\ref{fig5}a).

\subsection{Evaluation}
\label{subsec:sched_eval}

\begin{figure}[tb]
    \centering
    \setlength{\abovecaptionskip}{2pt}
    \includegraphics[width=0.9\linewidth]{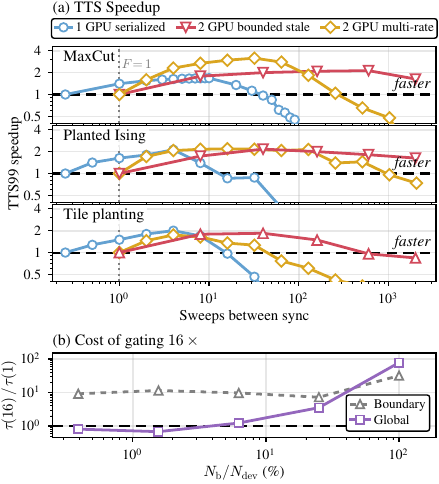}
    \caption[Evaluation provenance]{
    (a) Measured $\mathrm{TTS}_{99}$ speedup against synchronization interval, three kings families at $N=10^4$; the
    tile-planting two-device arm is the open-boundary sibling. (b) What gating the seam $16\times$ costs in mixing time.
    }
    \label{fig7}
\end{figure}

Fig.~\ref{fig7} puts the multi-rate, bounded-staleness, and
single-device schedules on one axis across three benchmark families:
MaxCut (Fig.~\ref{fig1}), frustrated loops, and tile
planting~\cite{perera2020chook,perera2020tileplanting}. We isolate the
multi-rate schedule's effect by withholding the boundary refresh, taking the
number of interior updates between refreshes as the synchronization interval for the multi-rate curve. We evaluate
the throughput--quality trade-off by sweeping the synchronization interval.

As barriers become less frequent,
$t_{\mathrm{sweep}}$ and $p_{\mathrm{succ}}$ both fall monotonically, and the TTS's interior optimum sits where the first stops outpacing the second (Eq.~\eqref{eq:tts}).

The blue curve varies $\gamma$ within a device; the yellow and red
curves vary $\gamma_{\mathrm{b}}$ across devices utilizing bounded staleness and multi-rate schedules. Within a single GPU
(blue), relaxing the barrier gains up to $2\times$ across all
benchmarks but degrades sharply beyond the optimum: every coupling in
the device is affected, and the barrier doubles as a cache-locality
synchronizer (Section~\ref{sec3}). Across devices (yellow, red), a
comparable synchronization interval yields up to $2.95\times$, and because only the
boundary couplings are affected, the benefit persists to much larger
$F$.

The interaction graph's degree and density set this $F$ tolerance: sparse graphs
often tolerate moderate staleness, while dense graphs require more frequent
synchronization because each stale update influences many
neighbors~\cite{pmlrv48sa16,aadit2022sparse}. We measure this on planted
frustrated-loop instances, hosted
on a degree-8 kings graph ($N{=}4{,}096$) and a degree-15 Pegasus graph
($N{=}5{,}640$). At 64 sweeps of staleness the residual energy above the planted
optimum rises by 5\% on kings against 30\% on Pegasus, each relative to its own
synchronous baseline, and at 256 sweeps by 14\% against 49\%. 

In a limited-restart million-spin study, synchronized, bounded-staleness, and multi-rate schedules achieve mean energy gaps of 0.72--2.19\% across three benchmark families~\cite{artifact2026}.

\subsection{Emulated Multi-Device Platform}
\label{subsec:emulation}

\begin{figure}[tb]
    \centering
    \setlength{\abovecaptionskip}{2pt}
    \includegraphics[width=0.9\linewidth]{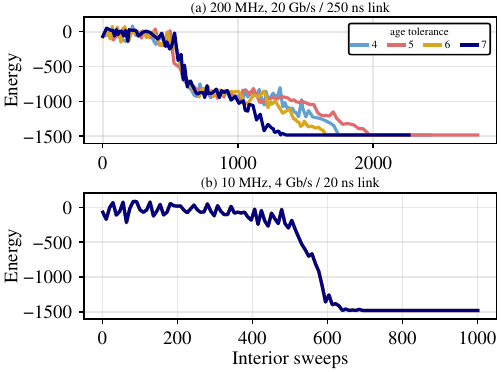}
\caption[Evaluation provenance]{Age tolerance on a $20\times20$ kings instance, energy against
interior sweeps, under FPGA emulation. (a) A 200\,MHz array on a $20$\,Gb/s, $250$\,ns link. (b) A 10\,MHz array on a $4$\,Gb/s, $20$\,ns link.}
    \label{fig10}
\end{figure}

The age-tolerance results of Fig.~\ref{fig10} run on an emulated machine rather
than on a board fleet. One large FPGA hosts several small p-bit array
instances, assembled under Vitis, and the inter-device link between them is not
a cable but a parameterized channel with a settable rate and a settable
latency. Moving the link inside the part is what makes the sweep possible: the
array clock, the link bandwidth, and the link latency become independent knobs. The PE arrays themselves are the synthesized design, with the same datapath, the same RNG, and the same update order as a deployed instance.

We evaluated multi-rate scheduling with bounded staleness
using a single FPGA to emulate a $20\times20$ kings graph partitioned across
four devices connected in a kings interconnect topology.

 Each spin carries an \emph{age}, the number of emulated clock cycles since its
value was last written by a local update or a link arrival, and a consumer
stalls rather than reading a spin past a configured \emph{age tolerance}, which
makes the tolerance $F_{\mathrm{b}}$ enforced per spin in clock cycles instead
of per sweep. Plotting energy against interior PE sweeps (Fig.~\ref{fig10})
normalizes out the interior rate, so the curves separate only when the boundary
lags, and whether they separate is set by the link rather than by the tolerance.
Over a low-latency link the boundary keeps up and every age tolerance traces one
curve; over a high-latency link the sweep budget to reach the ground state
spreads from ${\sim}1300$ to ${\sim}2000$ sweeps with no monotone ordering in the
tolerance. Multi-rate gating therefore pays only where the link cannot feed the
boundary, but on larger instances the loosest tolerances settle above the ground
energy, so the gain is bought against solution quality.

\section{Phase-Offset Streaming: Hiding $t_{\mathrm{comm}}$}
\label{subsec:mesh_sched}

  At small $N$, the $t_\mathrm{wave}$ term in
  Eq.~\eqref{eq:sweep_unified} is small, leaving $t_{\mathrm{comm}}$ exposed.
  The multi-rate and bounded-staleness schedules discussed in Section~V
  provide benefits in this regime. As $N$ grows, $t_\mathrm{wave}$ grows due to memory pressure. With only certain boundary waves depending on
  inter-chip transfers, as shown in Fig.~\ref{fig5}, this section presents a streaming dataflow to pipeline the transfer-dependent waves behind interior computation.

\begin{figure}[tbp]
    \centering
    \includegraphics[width=0.9\linewidth]{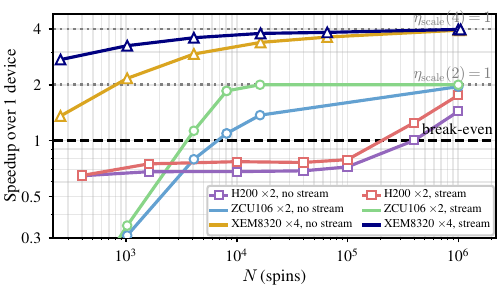}
    \caption{
 Two-device speedup and streaming speedup vs.\ problem size on two H200, two ZCU106 and four XEM8320. Ideal scaling efficiency for 2 devices and 4 devices is marked in grey dashed lines.
    }
    \label{fig6}
\end{figure}
\subsection{Algorithm and Benefits}
\label{sec:phaseoffset}

A sweep has $\chi W$ slots 
spread across $\varphi_{\mathrm{IT}}$ phases (Sections~\ref{subsec:exec_quantization} and~\ref{subsec:comm_term}),
and the discussion in this section assumes that the CG
matches the IT. An IT--CG mismatch can produce uneven
producer--consumer slack, limiting the flexibility to hide transfers through
scheduling. In practice, however, this effect is confined to a small set of spins: when Kings and Pegasus graphs are mapped onto a 2D-mesh IT, only 4 and 12 corner spins per chip, respectively, require two-hop routing, regardless of graph size. For these few multi-hop transfers, staleness can further relax the freshness requirement, complementing the techniques discussed in this section.

Each direction in a pair of devices in a maximal IT clique carries its own \emph{directed dependency}: $D_i$ produces its boundary spins, $D_j$ consumes them. The slack of the dependency $D_i \to D_j$ is the number of wave slots between $D_i$'s last producing slot and $D_j$'s first
consuming slot. The $\varphi_{\mathrm{IT}}$ devices of a maximal IT
clique cut a cycle of $\chi W$ wave slots into $\varphi_{\mathrm{IT}}$ gaps summing to $\chi W$, one gap of slack per
directed dependency. The phase-offset streaming algorithm designs the slot order $\sigma$ to maximize the minimum slack, which forces them equal at $\chi
W/\varphi_{\mathrm{IT}}$ slots (the half-sweep offset at $\varphi_{\mathrm{IT}}=2$).

 A schedule is \emph{latency-transparent} when:
\begin{equation}
\label{eq:hiding_eps}
t_{\mathrm{xfer}} \;\le\;
    \frac{t_{\mathrm{sweep}}}{\varphi_{\mathrm{IT}}}
    - t_{\mathrm{stage}},
\end{equation}
where $t_{\mathrm{stage}}$ is the staging time, from the start of a boundary spin's update to the moment its packet is assembled and ready to send over the link. $t_{\mathrm{stage}}$ should therefore be as small as possible, motivating smaller packets dispatched as soon as they are computed. If this equation's condition is satisfied, the transfer completes inside the latency-hiding window driving 
$\eta_{\mathrm{ovlp}}=1$, removing $t_{\mathrm{comm}}$ from Eq.~\eqref{eq:sweep_unified}.

Expanding $t_{\mathrm{xfer}}$ using Eq.~\eqref{eq:comm_unified} and writing $N_{\mathrm{dev}}=N/K$ and $\Theta_1 = N_{\mathrm{dev}}/t_{\mathrm{sweep}}$ ($\Theta_K$ being the $K$-device array rate) gives:

\begin{equation}
\label{eq:device_floor}
N \;\ge\; \varphi_{\mathrm{IT}}\,\Theta_K\,t_{\mathrm{lat}},
\qquad
\mathrm{BW}_{\mathrm{link}} \ge
   \frac{\varphi_{\mathrm{IT}}\, \Theta_K\, N_{\mathrm{b}}\,b_s}{N}. 
\end{equation}

The left equation shows the bandwidth-delay product in spins: a device must hold enough interior work to cover its own link's round trip. Fig.~\ref{fig6} shows this limitation together with when splitting work across two devices can be beneficial. 
For small $N$, streaming may only partially hide communication latency, leaving room for stale scheduling to improve performance (Section V). Larger $N_{\mathrm{dev}}$ widens the communication-overlap window and lowers the boundary fraction $N_{\mathrm{b}}/N_{\mathrm{dev}}$. The trade-off is between larger $K$ with smaller $N_{\mathrm{dev}}$, where staleness can mask exposed latency, and smaller $K$ with larger $N_{\mathrm{dev}}$, where more computation hides $t_{\mathrm{comm}}$. Eq.~\eqref{eq:hiding_eps}
estimates the exposed delay per boundary transfer, from which we can determine the staleness age $F$ needed to cover it. To evaluate this trade-off, we used a single FPGA to emulate a 400k-spin problem partitioned across either 16 or 32 devices. The 32-device configuration achieves $\mathrm{TTS}_{99}$ speedups of $3.5\times$ with a link configured at $\mathrm{BW}_{\mathrm{link}}=0.8$~Gb/s and $t_{\mathrm{lat}}=116$~ns, and $0.2\times$ when the link latency is changed to $300$~$\mu$s. Fig.~\ref{fig7} shows that for certain problems, large staleness can be tolerated before performance degrades, but this tolerance depends on the fraction and location of stale interactions.

 Fig.~\ref{fig6} reports speedup rather than TTS: under exact Gibbs, a $t_\mathrm{sweep}$ speedup translates directly into TTS reduction via Eq.~\eqref{eq:tts}. Also if we rearrange the equation with $\Theta_K$ on the left, we observe that only a larger problem or a lower-latency link buys more hideable throughput. For the equation on the right, for a geometrically local bounded-degree graph
that ratio is $O(1/\sqrt{N})$, so bandwidth demand falls away as devices grow and a correctly sized link is asymptotically latency-bound. Since on an all-to-all CG every spin is
a boundary spin, $N$ cancels. Also since $\varphi_{\mathrm{IT}} =K$, the requirement collapses to $\mathrm{BW}_{\mathrm{link}} \ge \Theta_K$, prohibiting scaling,
and reinforcing the scaling limitations of dense graphs.

\subsection{Mapping the PE Array onto the Partition}
\label{subsec:nesting}

Inside a device the partition is nested onto the $P_x \times P_y$ PE
array ($S = P_x P_y$, Fig.~\ref{fig5}c) in one of two ways. Mapping~A makes the \emph{wave} the
compact unit: each PE owns a strided sublattice, adjacent PEs touch
adjacent spins, every PE receives identical work each wave, and the
spin index map is a flat formula. Mapping~B makes the \emph{PE} the compact
unit: each PE owns a contiguous patch spanning all $W$ waves. The fraction of cross-patch couplings under Mapping~B falls as
$1/\sqrt{\ell_x \ell_y}$. Every such coupling is a non-local neighbor
access, and the platform sets what it costs: on a cache-based GPU it is
a scattered read against the memory hierarchy (the $t_\mathrm{mem}$
lever of Section~\ref{sec:memterm}); on an SP it is an inter-PE net or
a shared memory port. Mapping B shrinks both. It also localizes the inter-device link dependency: only partition-facing PEs stall
on a late halo, and under multi-rate scheduling the boundary PEs can
follow a link-determined rate while interior PEs run at full speed.
Mapping~A wins with balanced load, cheap
address generation, and coalesced access on a GPU: with matched storage layouts, Mapping~A is $1.20$ to $1.25\times$ faster across $N = 4\times10^6$--$3\times10^8$ on H200. Latency
hiding is indifferent to the mapping choice: with partition-facing waves scheduled
as in Section~\ref{sec:phaseoffset}, both mappings have the same latency-hiding window.
The FPGA and the fabricated chip in later sections adopt mapping~A with a color-minor order.

\subsection{2D Mesh}
\label{subsec:2dmesh}

\label{subsec:mesh_hw}
\begin{figure}[tbp]
    \centering
    \includegraphics[width=1\linewidth]{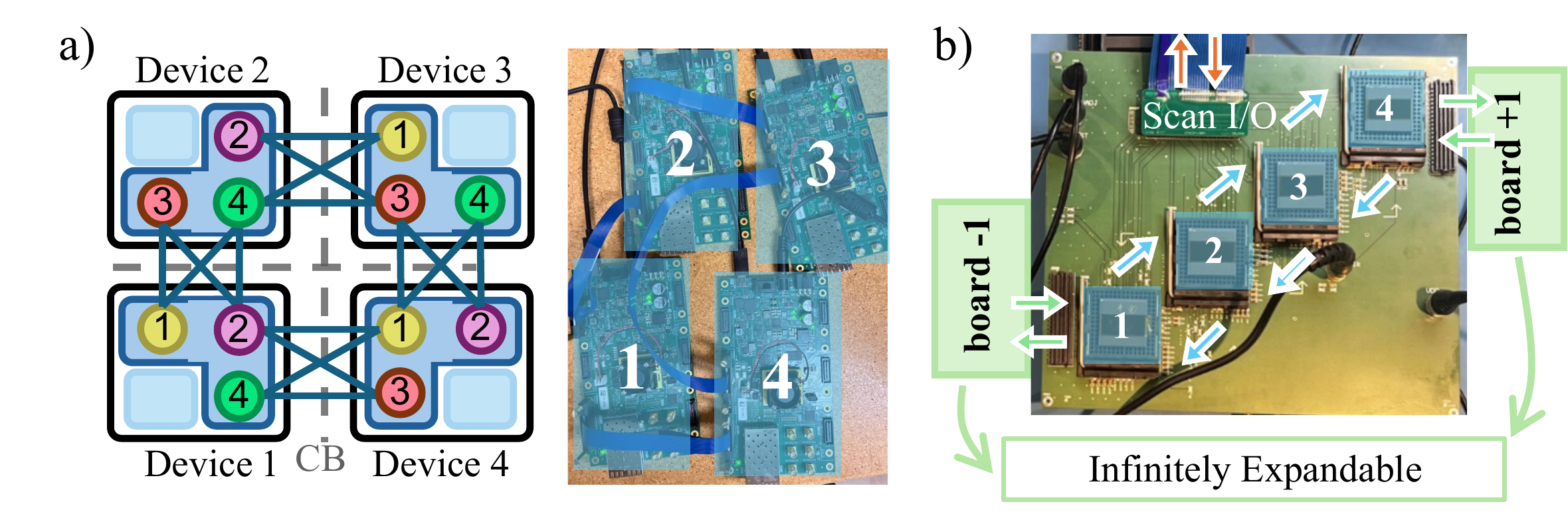}
    \caption{(a) 4 XEM8320 connected in a 2D mesh. (b) 4 ASICs connected in a 1D mesh.}
    \label{fig8}
\end{figure}

At the device level, the partition shape determines how the
boundary fraction scales with the number of devices. For a
$\sqrt{N}\times\sqrt{N}$ lattice partitioned into $K$ row slabs,
the per-device boundary fraction is $2K/\sqrt{N}$ and reaches
unity at $K=\sqrt{N}/2$. In contrast, a $\sqrt{K}\times\sqrt{K}$
block partition has boundary fraction $4\sqrt{K}/\sqrt{N}$,
preserving interior locality to larger $K$. We therefore use a
two-dimensional device mesh when scaling beyond the
two-device row-slab configuration.

To demonstrate this mapping, we connect four XEM8320 boards
as the $2\times2$ mesh shown in Fig.~\ref{fig8}. Each board
has one horizontal and one vertical nearest-neighbor link,
with a measured bandwidth of $806.4$\,Mb/s in each direction.
The board mesh matches the two-dimensional partition, so edge
halos are transferred in one hop; the two diagonal corner
couplings are forwarded through a neighboring board and
therefore require two hops. Depending on $t_\mathrm{sweep}$,
the communication may not be fully hidden: the link latency is $t_{\mathrm{lat}}=116$\,ns. Consequently, the transfer
latency remains on the critical path for small problem sizes, resulting in a sub-optimal scaling efficiency, shown in Fig.~\ref{fig6}.

\subsection{Multi-Chip ASIC}
\label{subsec:pegasus_asic}

The phase-offset streaming schedule is implemented on four ASICs
($1.5\,\mathrm{mm}^2$ each, TSMC 28\,nm)~\cite{wu2026pegasus}. We implement the Pegasus
graph ($d=15$~\cite{boothby2020pegasus}), chosen for its shorter
embedding chains~\cite{sajeeb2025scalable, iyer2025encoded}, at 6912 spins per chip.

Pegasus is not a planar lattice of spins, but it is a planar lattice
of Pegasus\emph{ cells}: each cell holds 24 spins spanning the graph's three orientations. Therefore, all
lattice dimensions in this section count
cells rather than spins. Each chip maps $L_x \times L_y = 12 \times
24$ cells onto a $3 \times 3$ array of \emph{Ising Units} (IU)
($P_x = P_y = 3$), each IU comprising 6 PEs that update one cell
together, so $S = 6\,P_x P_y = 54$ and the $32$ cells per IU
give $W = 32$ waves. All spins update in $\chi=4$ cycles at $\mathrm{II}=1$; couplings and
bias (Q6.3) sit in local per-IU SRAM.
Since the chip has no on-chip multiplier, the host loads
pre-scaled $\beta J$ and $\beta h$.

The streaming schedule moves inter-chip communication off the
critical path ($\eta_{\mathrm{ovlp}} \approx 1$): the chip's $\sigma$
places the north and south boundary rows at the start of each half of the 32-wave rotation, each dispatching to its neighbor while the array works through interior spins. In the $\chi W = 128$-slot rotation the packet dispatches as soon as
its boundary wave completes. The wave itself takes 4 cycles, and the
C2C interface runs at half the core clock: the 3-cycle asynchronous
FIFO crossing, counted in IO-clock cycles, charges 6 core cycles, so
$t_{\mathrm{stage}} = 4 + 6 = 10$ cycles. This leaves a 54-cycle
window or $415$\,ns of slack at the measured 130\,MHz clock.

With $\eta_{\mathrm{ovlp}} = 1$ and $t_{\mathrm{comm}} = 0$ in
Eq.~\eqref{eq:sweep_unified}, the schedule reaches peak throughput:
for $K=4$ ($N = 4 \cdot 6912 = 27{,}648$, $S = 4 \cdot 54 = 216$,
130\,MHz), Eq.~\eqref{eq:sweep_streaming} gives
$t_{\mathrm{sweep}} = 0.985\,\mu$s, equal to the single-chip sweep
time, for an aggregate $\Theta_4 = 28.1$\,Gflips/s (a numeric subscript on $\Theta$ counts devices). Adding chips
keeps this time fixed: a $K$-chip machine carries $54K$ PEs alongside
its $6912K$ spins, so the cycle count does not change with $K$.

However, maintaining this scaling behavior for larger chip
counts depends on the inter-chip routing and its ability to satisfy the
latency-hiding condition of Eq.~\eqref{eq:hiding_eps}.

Energy is $4.81$\,pJ per update measured at 130\,MHz with a 0.9\,V core supply and a 1.2\,V I/O supply.

 On one coherent
27k-node MaxCut problem across the on-PCB chip-to-chip links, with no
FPGA in the loop, the solution reaches $97.11\%$ of the global ground
energy; the annealed four-chip ladder on a 27k-node instance
reaches $99.91\%$ with silicon-resident state.

\section{Towards an Ideal Ising Accelerator}
\label{sec6}

\begin{table}[t]
\centering\footnotesize
\setlength{\tabcolsep}{3.5pt}

\caption{Performance gains from the three techniques}

\label{tab:levers}
\begin{tabular}{@{}lllc@{}}
\toprule
Technique & Control & Metric & Range \\
\midrule
Multi-rate    & refresh every sweep  & TTS$_{99}$           & $1.75$--$3.16\times$ \\
Bounded stale & $F_{\mathrm{b}}{=}1$ & TTS$_{99}$           & $1.84$--$2.17\times$ \\
Streaming     & same schedule in lock-step& $t_{\mathrm{sweep}}$ & $1.25$--$2.01\times$ \\
\bottomrule
\end{tabular}
\end{table}

\begin{figure}[tbp]
    \centering
    \includegraphics[width=0.9\linewidth]{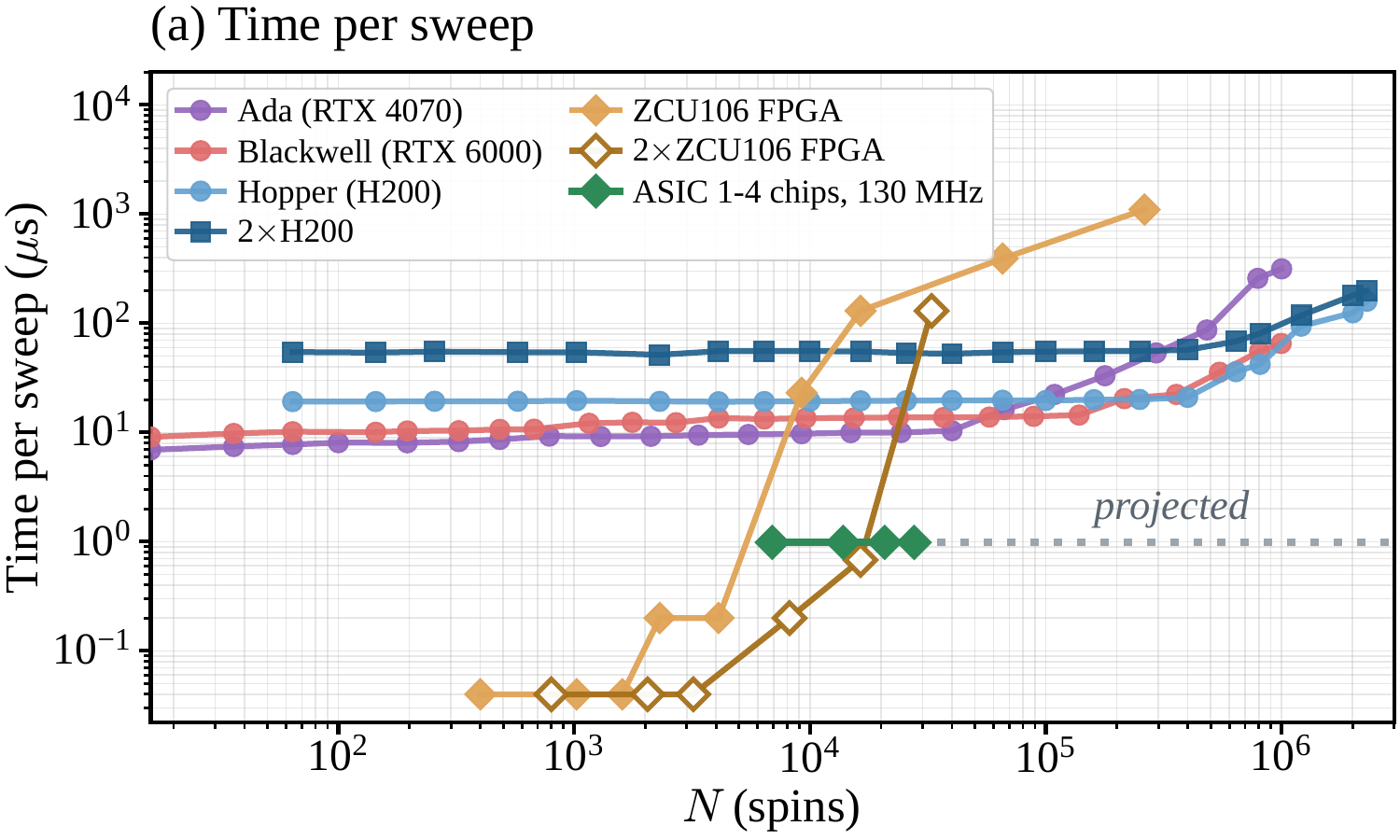}

    \vspace{0.25cm}

    \includegraphics[width=\linewidth]{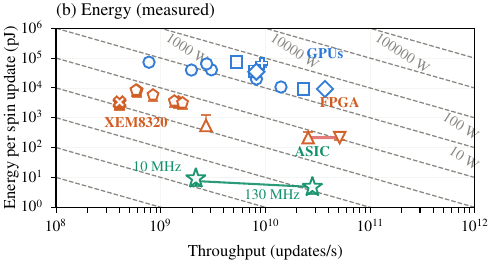}
    \caption[Evaluation provenance]{(a) Per-sweep time against $N$ for the platforms we measure,
    up to $10^6$ spins. Since the ASICs implement a streaming algorithm to hide
    $t_{\mathrm{comm}}$, adding more chips will keep $t_{\mathrm{sweep}}$ constant (extrapolated with dashed line). (b) Measured energy per spin update against sustained throughput; each point carries its accounting range as a vertical bar. Color is the platform class and marker shape the device and its count; the hollow markers and dotted bars are the XEM8320, whose power is a post-route estimate rather than a measurement.
    }
    \label{fig9}
\end{figure}

Table~\ref{tab:levers} collects the benefits for each of the three techniques presented in this paper. Fig.~\ref{fig9}a shows the per-sweep time each platform reaches. Fig.~\ref{fig9}b compares the platforms in measured energy per attempted spin update. 

The per-sweep time for the ASIC implementing the streaming dataflow is flat as $K$ increments. At $N{=}27{,}648$ the four-chip ASIC~\cite{wu2026pegasus} is
32.4$\times$ faster than the
topology-agnostic H200 CUDA kernel implementation used throughout this paper and 22.1$\times$ faster than a topology-specialized kernel optimized for the 27k-spin Pegasus graph through aggressive neighbor-index decoding and loop-unrolling.  However, this approach increases register pressure, reducing the number of resident warps
and limiting memory-level parallelism (MLP). This constrains the scalability of the memory-latency-bound
implementation. The general-purpose CUDA baseline used in this paper follows the approaches described in~\cite{whitehead2025pottsplayground,onizawa2024gpupsav,tarasiuk2026solvers3d}. At $N{=}27k$, the specialized H200 kernel achieves only
13\% occupancy. Concurrent execution of 29 independent chains increases occupancy to 74\%, at the cost of
a 56\% increase in $t_\mathrm{sweep}$, providing an opportunity to exploit applications such as adaptive parallel tempering~\cite{artifact2026}.

\leavevmode
Our model guides architectural design decisions
for a given Ising problem. 
At $N{=}16k$, partitioning the problem across two FPGAs yields a predicted local
computation time of 0.68~$\mu$s, compared with 130~$\mu$s
for the single-board DDR-based implementation. This $191\times$ reduction is consistent with our measurements and results from keeping the working set on-chip. When every partition fits on chip, our model predicts the crossover between communication-hidden and communication-exposed execution as the device count increases, discussed in Section~\ref{subsec:sched_eval}. For a given hardware configuration, the model also estimates the staleness age required to cover the exposed delay. The resulting effect on $\mathrm{TTS}_{99}$ depends on the trade-off between reduced execution time and any decrease in success probability.

The model can also predict cross-platform performance crossover points. The single-FPGA model predicts sweep times of $0.67$--$8.67$~$\mu$s, $1.28$--$16.7\times$ faster than the modeled RTX~4070 GPU results across eight King's graph sizes ($N=1\,\mathrm{k}$--$32\,\mathrm{k}$), consistent with measured results~\cite{artifact2026}. The GPU performance gap is primarily attributable to
architectural limitations, constrained
by memory latency and run-time dynamic scheduling. Our results suggest that statically scheduled PE arrays with local memory feeding an \(\mathrm{II}=1\) pipeline are a more promising architecture, exploiting the spatial concurrency and localized connectivity of sparse regular graphs while minimizing memory-latency
overhead. For GPU programmers, our findings highlight that coalesced access, warp residency, and sufficient occupancy to hide memory latency remain important considerations (Sections~III-E and~IV-B). Staleness is particularly beneficial on GPUs, where synchronization overhead and memory latency remain high even for small problem sizes. One promising direction for future work is to combine memory prefetching with bounded staleness, fetching the next subgraph’s couplings and biases while repeatedly updating the current subgraph using resident data to increase arithmetic intensity. Using our model, we predict that a Vera Rubin-class GPU needs $30.7\,\mu$s per sweep at $N{=}10^6$, $2.1\times$
faster than the measured H200 despite $4.6\times$ the memory bandwidth. These insights suggest a potential path for scaling toward million-spin Ising machines.

\section{Related Work}
\label{sec:relatedwork}

Our decomposition builds on BSP~\cite{valiant1990bsp},
LogP~\cite{culler1993logp}, and the roofline model~\cite{williams2009roofline}, including its variants~\cite{ilic2014cache,lopes2017gpuroofline,yang2020hierarchical,yang2020deeplearning,ding2019instruction}. Bounded staleness as a tunable knob comes from Hogwild~\cite{recht2011hogwild}
and SSP~\cite{ho2013ssp}, with convergence under asynchronous updates
established
theoretically~\cite{pmlrv48sa16,terenin2020async,burns2025parallel,onizawa2026unified}
and demonstrated on an all-to-all array by Snowball~\cite{hong2026snowball}. Aadit et
al.~\cite{aadit2026million} measure, per topology, the exchange-to-update ratio
above which networked FPGAs match a monolithic reference.

Ising accelerators differ in where couplings sit relative to update
logic, from resistive crossbars~\cite{afoakwa2021brim,sharma2022brim} and
near-memory digital
arrays~\cite{sundararaman2024sachi,chiang2024reaim,wu2026cobi,yang2024sophie,fu2025proca}
to lattice-coupled and fully connected
annealers~\cite{yamaoka2017,takemoto2019,aramon2019fujitsu,matsubara2020cmos,yamamoto2021statica},
partitioned simulated
bifurcation~\cite{goto2019sbm,goto2021classical,tatsumura2019fpga,tatsumura2021scaling,kashimata2023multichip},
and the deeply pipelined Janus family~\cite{baityjesi2012janus,baityjesi2014janusii,yoshimura2017,patel2020fpga,gyoten2018}.
Related dataflow accelerators keep the same neighbor-gather-then-local-update
structure for graph learning and mathematical
programming~\cite{song2024dsgl,wu2025dstpu,liu2025dstide,sundararaman2025spark,wang2024butterfly}.

Similar latency-hiding strategies have been used in stencil
and distributed scientific computing with the Message Passing Interface,
where boundary communication is overlapped with independent interior
computation~\cite{kim2012overlapping,pekkila2022scalable}.

Table~\ref{tab:ising_comparison} compares our implementations
against prior GPU-, FPGA-, and ASIC-based Ising implementations.

\begin{table}[!t]
\centering
\fontfamily{ptm}\fontsize{8}{9.5}\selectfont
\setlength{\tabcolsep}{1.5pt}
\renewcommand{\arraystretch}{1.05}
\caption[Comparison with prior Ising implementations]{Comparison with prior Ising implementations}
\label{tab:ising_comparison}
\begin{tabular*}{\linewidth}{@{\extracolsep{\fill}}llrrr@{}}
\toprule
\textbf{Work} & \textbf{Platform} & $\boldsymbol{N}$ &
\shortstack{\textbf{Gflips/s}} &
\shortstack{\textbf{nJ/flip}} \\
\midrule
\multirow{4}{*}{Ours$^{\dagger}$} & 2$\times$ H200 & 4.2M & 14.3 & 60.8 \\
& 4$\times$ 28 nm ASIC~\cite{wu2026pegasus} & 27.6k & 28.1 & 0.00481 \\
& 4$\times$ ZCU106 & 32.8k & 0.697 & 15.5 \\
& 4$\times$ XEM8320 & 4k & 1.61 & 61.3 \\
\midrule
Yang '19~\cite{yang2019ising} & V100 & N/R & 11.37 & 21.99 nJ$^{*}$ \\
Chowdhury et al. '25~\cite{chowdhury2025pushing} & 1$\times$ U250 & 4k & 185 & 50 pJ \\
Kim--Sim '25~\cite{kim2025ising} & 4$\times$ 28 nm ASIC & 32.8k & 2.64$^{*}$ & 73.3 pJ$^{*}$ \\
COBI '26~\cite{wu2026cobi} & 65 nm ASIC & 64 & 0.889$^{*}$ & 0.585 nJ$^{*}$ \\
Aadit '26~\cite{aadit2026million} & 18$\times$ VP1902 & 1M & 1,000 & N/R \\
\bottomrule
\end{tabular*}

\smallskip
\begin{minipage}{\linewidth}
\fontsize{8}{9.5}\selectfont
$^{*}$ Derived\quad
$^{\dagger}$ Our entries show each platform's best reported throughput.
\end{minipage}
\end{table}

\section{Conclusion}

We provide the first calibrated cross-platform performance model
for distributed digital Ising machines. It extends the
roofline analysis by explicitly parameterizing the execution structure of the
Gibbs sampling algorithm, yielding the following design guidance that was previously unclear to the community:
(i) residency-driven data placement, (ii) bounded-staleness
and multi-rate scheduling, and (iii) a phase-offset streaming dataflow.

We achieve up to (i) 4.2$\times$ speedup, (ii) 3.16$\times$/2.17$\times$ lower \TTS, and (iii) 2.01$\times$ speedup. A
22.1$\times$ speedup was also measured comparing a topology-specialized H200 baseline against a fabricated four-chip 28\,nm ASIC~\cite{wu2026pegasus} that sweeps 27,648 spins in 0.985\,$\mu$s at 4.81\,pJ per update.

\section*{Acknowledgements}
This work was supported by the Carnegie Mellon University Department of Electrical and Computer Engineering (CMU ECE). The authors acknowledge the CMU ECE Computing Research GPU Cluster for providing access to NVIDIA H100 and H200 systems used in this study.

The Streaming ASIC prototype and multi-chip platform were developed with support from CMU ECE research infrastructure and the Apple NSI program. The authors thank Apple for its support.

\nocite{pervaiz2019weighted,aadit2022sparse,baityjesi2014janusii,aadit2026million}

\clearpage
\appendix[AI Use]

The authors used AI assistants, primarily Codex (ChatGPT),
in preparing this paper. The tools contributed to drafting and copy-editing
the prose and to writing the supporting code that generates the workloads,
processes the data, and produces the figures. To a lesser degree they helped
sanity-check derivations and summarize related work. The analytical model, methodology,
measurements, and conclusions are the authors' own. The authors reviewed,
tested, and verified all AI-assisted output and take full responsibility for
the content and accuracy of this work.

\end{document}